\documentclass{cimento}
\usepackage{graphicx}
\usepackage{amssymb}
\date{}
\usepackage{amsmath}

\title{Spherical Crystallography: Virus Buckling and Grain Boundary Scars}

\author{David R. Nelson}

\instlist{\inst{} Lyman Laboratory of Physics, Harvard University, Cambridge, MA 02138}

\begin{document}

\maketitle
\thispagestyle{empty}

\begin{abstract}
Ordered states on spheres require a minimum number of topological defects. For the case of 
crystalline order, triangular lattices must be interrupted by an array of at least 12 five-fold
disclination defects,
typically sitting at the vertices of an icosahedron. For $R>>a$, where $R$ is the sphere radius
and $a$ the particle spacing, the energy associated with these defects is very large.
This energy can be  lowered, however,  either by buckling, as appears to be the case for 
large viruses, or by introducing unusual finite length grain 
boundary scars. The latter have been
observed recently for colloidal particles adsorbed onto water droplets in oil. 
\end{abstract}
\maketitle


\section{Introduction}

Understanding ordered states on curved surfaces like the sphere requires dealing with 
``geometrical frustration.'' Order parameters which describe crystalline or hexatic
order in the centers of mass of the microscopic degrees of freedom, or the projections
of tilted molecules onto a local tangent plane, often minimize their energy by aligning
in neighboring regions. It is natural to assume that this alignment can be achieved by 
parallel transport of the vector or tensor order parameter which represents the local 
order [1, 2]. In a continuum description, the required parallel transport can be described
by connection coefficients which couple to the displacement or phase angle much like a vector
potential couples to the phase of a Type II superconductor [3]. However, such alignment 
cannot be achieved everywhere when a nonzero Gaussian curvature is present. In 
particular, an attempt to produce a vector field aligned by parallel transport fails on a loop
which contains a net integrated Gaussian curvature [4]. The Gaussian curvature embodied in 
the line integral of the connection coefficients around the loop resembles the frustration 
in a superconducting order parameter produced by a nonzero magnetic field [5].

Consider  $N$ particles with a repulsive hard core interaction closed-packed  on the 
surface of a sphere. If $R$ is the radius of sphere, $N\sim(R/a)^2$, where $a$ is the hard core diameter.
Now imagine ``triangulating'' this sphere by connecting nearest neighbor particles 
using the spherical analogue of the familiar Wigner-Seitz construction of solid state 
physics. The geometric frustration discussed above appears on a more microscopic level
in Euler's theorem [6] on a sphere, which states the number of triangles $T$, edges $E$ and 
vertices $N$ of such a tesselation are related,
\begin{equation}
T-E+N=2.
\end{equation}
Each edge is shared with two triangles, but a single triangle has three edges, so 
$T=\frac{1}{3}(2E)$ and Eq. (1) can be rewritten as $N-\frac{1}{3}E=2$. 
If the coordination numbers at each node in this triangulation are denoted $Z_j$, then 
$E=\frac{1}{2}\sum_j Z_j$ and 
Euler's theorem constrains the deviations of the $\{Z_j\}$ from 6 on any surface with the topology of
a sphere,
\begin{equation}
\sum_{j=1}^N (6-Z_j)=12.
\end{equation}
Although identical particles with a hard core repulsion in a plane typically pack into a 
triangular lattice with six-fold coordination, on a sphere Eq. (2) shows that there must  
be at least 12 five-fold disclinations as well as an arbitrary 
number of six-coordinated sites. The 12 extra disclination defects
(each representing a rotational discontinuity of $2\pi/6$) can be viewed as a consequence
of the geometrical frustration associated with the Gaussian curvature  of a sphere, which
integrates to $4\pi=12\times (2\pi/6)$. One plausible candidate for the ground state of an
array of particles on the sphere is a ``superlattice'' of these disclinations (rather like the 
Abrikosov flux lattice of a Type II superconductor) arrayed at the vertices of a regular 
icosahedron.

Although this hypothesis is likely to be approximately correct for the ground state of 
$N$ interacting particles on a sphere for moderate values of $R/a$, the very large
energy associated with the twelve isolated disclinations makes this guess questionable
when $R/a$ becomes large. In fact, the extra energy  due to the disclinations (relative 
to a perfect triangular lattice in flat space) grows like $YR^2$ (see Sec. 2), where $Y$ is a 
two-dimensional Young's modulus. In this review, we briefly sketch two ways of 
reducing this energy. The first, relevant to  hollow spherical shells which can be 
used to model viruses, invokes a buckling transition of the 12 disclinations and leads to 
a \emph{faceted} icosahedral shape [7]. If $\kappa$ is bending rigidity of the spherical
shell, the $YR^2$ behavior of the disclination energies is replaced by $\kappa\ln  (R/a)$ 
for large $R$. Another mechanism takes over for particles packed on a large spherical 
droplet, where surface tension limits significant buckling of disclinations out the local tangent 
plane.Even in this case, the energy can still be reduced by introducing grain boundaries 
which radiate from the 12 disclinations [8]. The energy now grows like $E_c(R/a)$, where
$E_c$ is a dislocation core energy. This energy is larger than if buckling of the disclinations
were allowed, but still smaller than the amount $\sim YR^2$ characteristic of isolated
disclinations when $R$ is large..

In Sec. 2, we review virus buckling, while Sec. 3 presents a  brief discussion of
grain boundary ``scars'' which have now been observed experimentally [9].   
\bigskip
\bigskip

\section{Theory of Virus Shapes}

Understanding virus structures is a rich and challenging problem [10], with a wealth of new 
information now becoming available.  Although traditional X-ray crystallography still 
allows the most detailed analysis [11], three-dimensional reconstructions of icosahedral 
viruses from cryo-electron micrographs are now becoming routine [12].  
Electron microscope images of many identical viruses in a variety of 
orientations are used to reconstruct a three-dimensional image on a computer, similar 
to CT (computed tomography) scans in medical imaging. There are now, in addition, 
beautiful single molecule experiments which measure the work needed to load a virus 
(bacteriophage $\phi$29) with its DNA package [13]. It is of some interest 
to explore the elastic parameters and physical ideas which determine the shapes of 
viruses with an icosahedral symmetry.  

The  analysis of approximately spherical viruses dates back to pioneering work by 
Crick and Watson in 1956 [14],  who argued that the small size of the viral genome 
requires identical structural units packed together with an icosahedral symmetry.   These 
principles were put on a firm basis by Don Caspar and Aaron Klug in 1962 [15], who 
showed how the proteins in a viral shell (the ``capsid'') could be viewed as icosadeltahedral 
triangulations of the sphere by a set of pentavalent and hexavalent  morphological units 
(``capsomers'').  The viral shells (there can also be an outer envelope composed of  
membrane elements from the host cell) are characterized by a pair of integers $(P,Q)$  such 
that the number of morphological units is $N = 10(P^2 + PQ + Q^2) + 2$. To get from one 
pentavalent capsomer to another, one moves $P$ capsomers along a row of near-neighbor 
bonds on the sphere, turns 60 degrees and moves another $Q$ steps. The Euler theorem 
discussed in the Introduction insures that 
the number of capsomers in  five-fold environments is exactly 12.   A simple icosahedron 
of 12 morphological units corresponds to (1,0) while soccer balls and C$_{60}$ fullerene  
molecules are (1,1) structures with 32 polygons. An icosadeltahedron with symmetry 
indices (3,1) is shown in Fig. 1. The polyoma virus (SV40) is a (2,1) 
structure with 72 capsomeres, while the much larger adenovirus and herpes simplex virus 
are (5,0) and (4,0) structures with 252 and 162 morphological units, respectively.    
Structures like that  in Fig. 1 with $P$ and $Q$ nonzero and $P \not=  Q$ are chiral.

\begin{figure}[htbp]
\begin{center}
\includegraphics{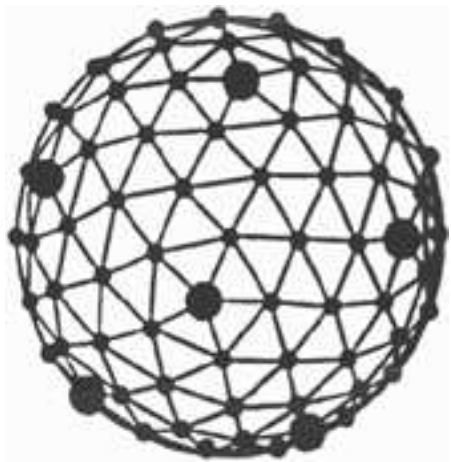}
\caption{{A right-handed (3,1) triangulated net 
(icosadeltahedron) used to describe virus structure.
The (1,3) structure is left-handed.}}
\end{center}
\end{figure}

Note that the relatively small polyoma virus (diameter 440 Angstroms) is round (see Fig. 
2a), while the much larger herpes simplex virus (diameter 1450 Angstroms) has a more 
angular or faceted shape [16] (see Fig. 2b).  Faceting of large viruses is in fact a common 
phenomenon; the protein subunits of different viruses, moreover,  have 
approximately the same molecular weight [17]. If  these protein assemblies are 
characterized by elastic constants and a bending rigidity [18], it is interesting to  ask how 
deviations from a spherical shape develop with increasing  virus size.      

\begin{figure}[htbp]
\begin{center}
\includegraphics{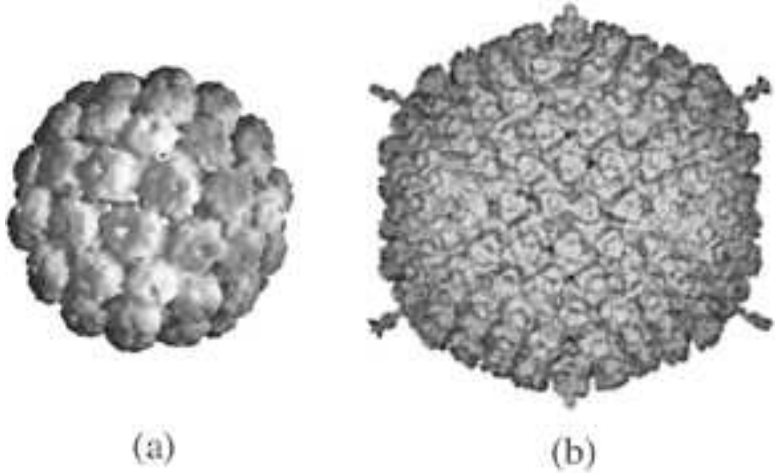}
\caption{{The polyoma virus (a) is approximately spherical while the larger
adenovirus (b) is more faceted (not to scale) [16].
}}
\end{center}
\end{figure}

The faceting of large viruses may in fact  
be caused by a buckling transition associated 
with the 12 isolated points of 5-fold symmetry.   These 
singularities can be viewed as disclinations in an otherwise 6-coordinated medium.    It is 
well known that the large strains associated with an isolated disclination in a {\it flat} 
disk of triangular 
lattice leads to buckling into a conical shape for [1, 19]
\begin{equation}
YR^2/\kappa\geq 154,
\end{equation}
where $Y$ is the two-dimensional Young's modulus, $\kappa$ is the bending 
rigidity and  $R$ is the 
disk radius. The energy of a single 5-fold disclination with ``charge'' $s = 2\pi/6$ 
centered in a flat array of proteins of size $R$ is approximately 
\begin{equation}
E_5\approx {1\over 32\pi} s^2YR^2.
\end{equation}
However, above a critical buckling radius $R_b\approx\sqrt{154\kappa/Y}$, is there
a conical deformation  with [19]
\begin{equation}
E_5\approx (\pi/3)  \kappa \ln 
(R/R_b)+ {1\over 32\pi} s^2YR_b^2.
\end{equation}
One might expect a similar phenomenon for 12 disclinations confined to a surface with a 
spherical topology.   Indeed, the elastic energy for 12 disclinations on an undeformed 
sphere of radius $R$ is has a form similar to Eq. (4), namely [8]
\begin{equation}
E \approx  0.604(\pi Y R^2/36),
\end{equation}
where the sphere radius $R$ now plays the role of the system size.  Although it 
seems highly likely that these 12 disclinations can lower their energy by buckling for 
large $R$, the nonlinear nature of the underlying elastic theory [18] leads to complex 
interactions between the resulting conical deformations. A boundary layer analysis of 
the von Karman equations for bent plates predicts anomalous scaling for the mean 
curvature in the vicinity of the ridges connecting conical singularities [20]. Interesting 
scaling behavior also arises in the vicinity of the apexes of the 
cones themselves [21].     
Another  intriguing physical realization of the buckling problem 
(with $R/a\gg 1$ so that  boundary layers near ridges are important) lies in the faceting of 
lecithin vesicles at temperatures sufficiently low so that the lipid constituents have 
crystallized [22].

To search for a  buckling transition in viral capsids or 
crystalline vesicles, one must study the strains and ground states of crystalline 
particle arrays with 12 disclinations in a spherical geometry [7]. The nonlinear 
Foppl-von Karman 
equations for thin shells with elasticity and a bending energy can be treated using a 
floating mesh discretization developed and studied extensively in the context of ``tethered 
surface'' models of polymerized membranes [23].   By taking the nodes of the mesh to 
coincide with the  capsomers, even small viruses can be handled in 
this way, although any buckling transition will surely be smeared out unless $R_b/a$ is large, 
where $a$ is the spacing between morphological units.  Ideas from continuum elastic theory 
will, of course, be most applicable for vesicles composed of many lipids and for large 
viruses --- Viruses with as many as 1692 morphological units have been reported [24].   

There may be  inherent 
limitations on the size of viral capsids that follow from the elastic properties of thin 
shells.  Because larger viruses can accommodate more genetic material, larger sizes could 
confer an evolutionary advantage. If, however, large viruses buckle away from a 
spherical shape, the resistance of the capsid to mechanical deformation will degrade. 
A theory of buckled crystalline order on spheres  
allows estimates of important elastic parameters of the capsid shell from structural 
data on the shape anisotropy [7].   Although  some aspects  of virus structure could be accounted for 
by the physics of shell theory, other  features  could be more relevant  
to cell recognition, avoidance of immune response, etc.
Estimates of  quantities such as the bending rigidity and 
Young's modulus of  a empty viral shell might allow an understanding of  
deformations due to 
loading with DNA or RNA [13].

The results  of recent  investigations of buckling 
transitions in a spherical geometry using the methods of  Refs. [19] and [23] are
illustrated in Fig. 3 [7].  Spherical shells  
do indeed become aspherical as the ``von Karman number'' 
$YR^2/\kappa$ increases from values of order unity to $YR^2/\kappa\gg  1$.   
The 
mean square ``asphericity'' (deviation from a perfect 
spherical shape) departs significantly 
from zero when $YR^2/\kappa$  exceeds 154, the location of the buckling transition  in flat 
space.  Note that fits of buckled viruses or crystalline vesicles to this universal 
curve would allow an experimental estimate of the ratio $Y/\kappa$.  
More quantitative information on the buckled shape can be extracted  by expanding 
the radius $R(\theta,\phi)$ in spherical harmonics,
\begin{equation}
R(\theta,\phi)=\sum_{\ell=0}^\infty
\sum_{m=-\ell}^\ell
Q_{\ell m}
Y_{\ell m}
(\theta,\phi),
\end{equation}
and studying the  rotationally invarient quadratic invariants allowed for icosahedral viruses 
or vesicles, namely 
\begin{equation}
\hat{Q}_\ell=
\sqrt{\frac{1}{2\ell+1}
\sum_{m=-\ell}^\ell|Q_{\ell m}|^2} /Q_{00}
\end{equation}
with $\ell= 0, 6, 10, 12, 16, 18, \cdots$ [25].  Although {\it any} parameter set  of the form
$\{\hat{Q}_6, 
\hat{Q}_{10}, \hat{Q}_{12}\cdots\}$  could be consistent with an icosahedral symmetry, all 
buckled objects describable by  the theory of 
elastic shells should in fact  lie on a universal curve 
parametrized by the value of  $YR^2/\kappa$.   
Deviations from this curve are presumably be 
of some biological interest. In addition to exploring shape changes induced by an 
internal pressure (A virus loaded with 6 $\mu$m of DNA can have an internal pressure as 
large as 6 MPa! [13]), it would interesting  to study  the spontaneous membrane curvature 
produced by an asymmetric conical shape of the constituent proteins [26].  
See Ref. [7] for more details and quantitative fits to the shapes of the HK97 virus
and a yeast virus.
\begin{figure}
\begin{center}
\includegraphics{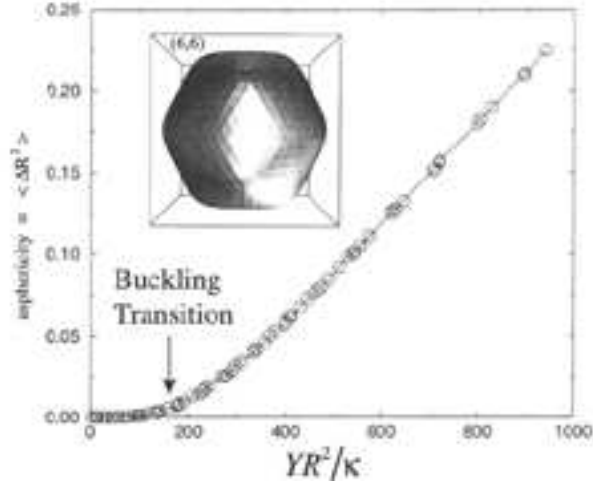}
{\caption{Mean square asphericity as a function of YR$^2/\kappa$ for many 
different icosadeltahedra. The inset shows a (6,6) structure with 
YR$^2/\kappa\approx 1000$.}}
\end{center}
\end{figure}

\section{Grain Boundary Scars}

When particles are packed on a surface where the restoring force is a surface tension,
instead of a bending rigidity, buckling is suppressed. For example, water droplets 
in oil coated with colloidal particles remain spherical even in the presence of 
disclinations in the colloidal array. 
Remarkably, the classical ground state of interacting particles confined to a 
frozen surface like a sphere is still a matter of debate [27],  especially in the limit of large 
sphere radius $R$ compared to the particle spacing $a$. This problem was first posed for 
interacting electrons by J. J. Thomson in 1904 [28].  Physical realizations 
include multi-electron
bubbles in superfluid helium [29],  and ``colloidosomes'' [30], which 
are lipid bilayers or droplets coated with 
colloidal particles. Colloidosomes are 
potential delivery vehicles for drugs, flavors and fragrances.  
To study the ground states of such systems, one can analyze
an effective free energy describing the physics of disclination arrays
constrained to lie on an arbitrary two-dimensional surface. 
For spherical surfaces,  the ground state for 
moderate $R/a$ can be approximated by a triangular lattice with twelve 5-fold disclinations 
at the vertices of an icosahedron. Building on work by  Alar Toomre [31] and by 
Michael Moore and collaborators [32], it is possible to construct a detailed theory of interacting
defects on a frozen, spherical topography [8]. Although the buckling found in viruses is not 
allowed, the ground state for large $R/a$ 
includes unusual finite length grain boundaries. 
The  methods of Ref. [8]  allow the direct minimization of 
the energy of, say, 26,000 interacting particles to be replaced by the much easier problem 
of  finding the ground state of  132  interacting 
disclination defects, [see Fig. 4] with interactions 
determined by continuum elastic theory on the sphere. 
Excellent results can be obtained using the Young's modulus and 
the dislocation core energy of particles in flat space as 
input parameters. Work by Bausch \emph{et al.} [9] on  
colloidal particles absorbed onto water droplets in oil indicates a conventional
ground state with just 12 disclination defects for small $R/a$, but with 
extra dislocations associated with grain boundaries for $R/a\ge 5$, corresponding to 
particle numbers $N\gtrsim N _c\approx 360$. Both this critical value of $N$  and the 
number of excess dislocations in the grain boundary arms for $N>N_c$ can be estimated using the 
theory of Ref. [8], with good agreement with experiment.
\begin{figure}[h]
\begin{center}
\includegraphics{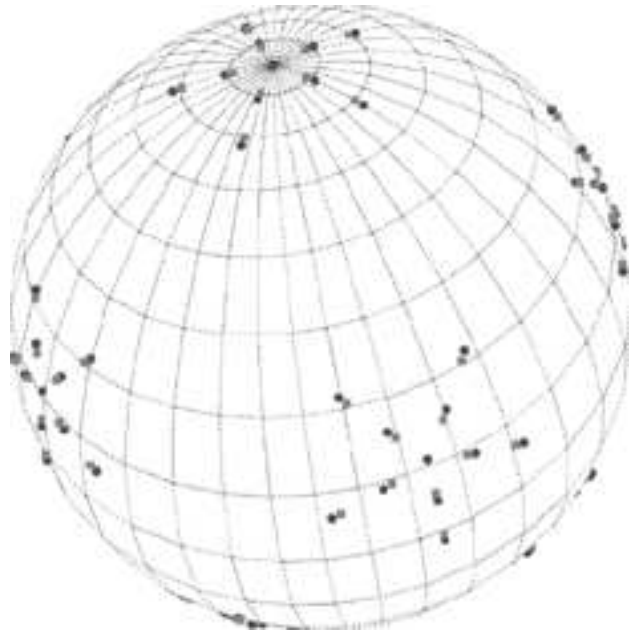}
{\caption{Dislocations emerging from 5-fold disclinations
(circles) for a crystal of aproximately 26,000 particles on a 
sphere [4]. Each dislocation is a 5--7 (circle-square) disclination
{\it pair} in this representation.
}
}
\end{center}
\end{figure}

See Ref. [33] for a discussion of the energetics of the crystalline state for 
large particle numbers and power-law potentials when \emph{both} 
buckling and grain boundary scars are forbidden.
\bigskip
\bigskip

\acknowledgments

The work described here was carried out in collaboration with M. Bowick, 
J. Lidmar, L. Mirny and A. Travesset. It is a pleasure to acknowledge these fruitful 
collaborations, as well as discussions with S. Harrison and A. Toomre. This work was
supported by the National Science Foundation, through Grant DMR-0231631 and 
in part through the Harvard Materials Research Laboratory, via Grant DMR-0213805.


\end{document}